\documentclass{egpubl}
\usepackage{eg2025}
\ConferencePaper
\CGFccby

\usepackage{times}
\usepackage[T1]{fontenc}
\usepackage{dfadobe}

\usepackage{cite}

\BibtexOrBiblatex

\electronicVersion
\PrintedOrElectronic

\ifpdf \usepackage[pdftex]{graphicx} \pdfcompresslevel=9
\else \usepackage[dvips]{graphicx} \fi

\usepackage{egweblnk}

\usepackage{lineno}
\usepackage{amssymb}

\ifpdf \usepackage[pdftex]{graphicx} \pdfcompresslevel=9
\else \usepackage[dvips]{graphicx} \fi

\usepackage{egweblnk} 
\usepackage{booktabs}
\usepackage{graphicx}
\usepackage{enumitem}
\usepackage{soul}
\usepackage{dsfont}
\usepackage{gensymb}
\usepackage{microtype}
\usepackage{pifont}
\usepackage{threeparttable}
\usepackage{collect}
\usepackage{xspace}
\usepackage{xfrac}
\usepackage{array}
\usepackage{multirow}
\usepackage{wrapfig}
\usepackage{url}
\usepackage[nolist]{acronym}
\usepackage{amsmath} 
\usepackage{pifont}
\newcommand{\cmark}{\checkmark}%
\newcommand{\xmark}{\scalebox{0.85}{\ding{53}}}%

\DeclareGraphicsExtensions{.png,.jpg,.pdf,.ai,.psd}
\def\figurePath{figures/}

\usepackage{xr}
\makeatletter
\newcommand*{\addFileDependency}[1]{
  \typeout{(#1)}
  \@addtofilelist{#1}
  \IfFileExists{#1}{}{\typeout{No file #1.}}
}
\makeatother

\newcommand{\refFig}[1]{Fig.~\ref{fig:#1}}

\newcommand{\refTab}[1]{Tab.~\ref{tab:#1}}

\newcommand{\refSec}[1]{Sec.~\ref{sec:#1}}

\newcommand{\refEq}[1]{Eq.~\ref{eq:#1}}

\usepackage[bold=0.55]{xfakebold}

\usepackage{icomma}

\newcommand{\myfigure}[2]{%
    \begin{figure}[htb]%
    \centering\includegraphics*[width = \linewidth]{\figurePath#1}%
    \vspace{-.2cm}%
    \caption{#2}%
    \vspace{-.5cm}%
    \label{fig:#1}%
    \end{figure}%
}

\newcommand{\mycfigure}[2]{%
    \begin{figure*}[htb]%
    \centering\includegraphics*[width = \linewidth]{\figurePath#1}%
    \vspace{-.1cm}%
    \caption{#2}%
    \vspace{-.7cm}%
    \label{fig:#1}%
    \end{figure*}%
}

\soulregister\ref7
\soulregister\cite7
\soulregister\refFig7
\soulregister\refAlg7
\soulregister\ref7
\soulregister\pageref7
\soulregister\shortcite7
\soulregister\eg0
\soulregister\ie0
\soulregister\etal0

\DeclareGraphicsExtensions{.png,.jpg,.pdf,.ai,.psd}
\newcommand{\mysection}[2]{\section{#1}\label{sec:#2}}
\newcommand{\mysubsection}[2]{\subsection{#1}\label{sec:#2}}

\DeclareRobustCommand{\change}[1]{{#1}}

\newcommand{\eg}{e.g.,\ }
\newcommand{\ie}{i.e.,\ }
\newcommand{\etal}{et~al.\ }

\newcommand{\mymath}[2]{
    \newcommand{#1}{\TextOrMath{$#2$\xspace}{#2}}
}

\begin{acronym}
\acro{DDPM}{denoising diffusion probabilistic model}
\acro{HDR}{high-dynamic range}
\acro{LDR}{low-dynamic range}
\acro{TMO}{tone mapping operator}
\acro{DPS}{diffusion posterior sampling}
\end{acronym}

\usepackage{arydshln} 

\title{Bracket Diffusion:\\HDR Image Generation by Consistent LDR Denoising}

\author[Bemana et al.]{
\large{
    Mojtaba Bemana\textsuperscript{1} \hspace{5mm}
    Thomas Leimk\"uhler\textsuperscript{1} \hspace{5mm}
    Karol Myszkowski\textsuperscript{1} \hspace{5mm}}
    \\
    \large{Hans-Peter Seidel\textsuperscript{1} \hspace{5mm}
     Tobias Ritschel\textsuperscript{2} \hspace{5mm}
    } 
    \\
    \\
    \textsuperscript{1}Max-Planck-Institut f\" ur Informatik \hspace{5mm}
    \textsuperscript{2} University College London
}

\begin{document}

\teaser{
   \vspace{-0.9cm}
    \includegraphics[width=0.95\linewidth]{figures/Teaser}
    \centering
    \caption{
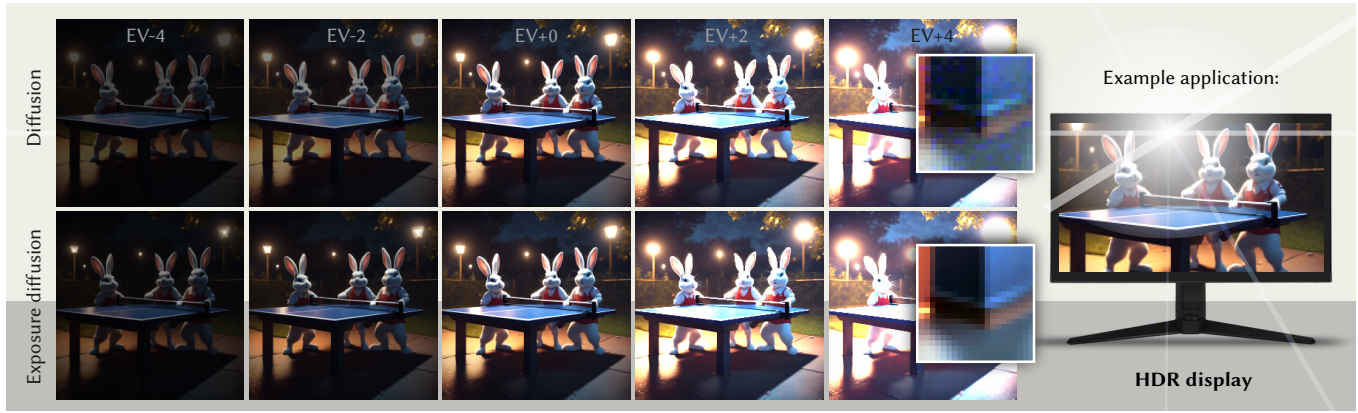
 Existing denoising diffusion models (top row) generate images with \acf{LDR} on a certain exposure in the center.
 When re-exposed to other levels, bright parts like the lamps do not retain their contrast, and dark areas do not reveal details as in the shadow below the table.
 In our \acf{HDR} approach (bottom), diffusion is performed at multiple exposure brackets, such that the lamps retain their contrast and the details in the animals' bodies are produced without noise (see insets).
 An example application is an \ac{HDR} display, where high pixel values map to high physical intensity.
    }\label{fig:Teaser}
   }
   
\maketitle

\begin{abstract}
We demonstrate generating \ac{HDR} images using the concerted action of multiple black-box, pre-trained \ac{LDR} image diffusion models.
Common diffusion models are not \ac{HDR} as, first, there is no sufficiently large \ac{HDR} image dataset available to re-train them, and, second, even if it was, re-training such models is impossible for most compute budgets.
Instead, we seek inspiration from the \ac{HDR} image capture literature that traditionally fuses sets of \ac{LDR} images, called ``exposure brackets'', to produce a single \ac{HDR} image.
We operate multiple denoising processes to generate multiple \ac{LDR} brackets that together form a valid \ac{HDR} result.
To this end, we introduce a brackets consistency term into the diffusion process to couple the brackets such that they agree across the exposure range they share.
We demonstrate \ac{HDR} versions of state-of-the-art unconditional and conditional as well as restoration-type (LDR2HDR) generative modeling.


 




\end{abstract}



\mysection{Introduction}{Introduction}
Images generated by modern denoising diffusion models \cite{rombach2022high-resolution,sohl2015deep} have shown an unprecedented combination of user control and image quality.
Unfortunately, the resulting images are \ac{LDR} while in computer graphics, several applications, such as physically-based simulation and rendering \change{\cite{debevec1998rendering,Reinhard2010HDR}}, scene reconstruction with significant shadows and specular highlights \cite{jun2022hdr,huang2022hdr,mildenhall2022nerf}, as well as \change{advanced television displays \cite{lei2024mini,shimazu20113d,seetzen2004high}, and emerging virtual reality systems \cite{zhong2021reproducing,zhao2020high}, rely on the capabilities of {\ac{HDR}} imaging.}

We propose to close this gap by introducing a simple and effective method to upgrade a black-box denoising diffusion model from \ac{LDR} to \ac{HDR} image generation.

This poses two main challenges: first, the \emph{limited scale of the available HDR training data}, which is orders of magnitude lower than its LDR counterpart, and second, the fact that for most users, it is \emph{impossible to re-train the denoiser due to the sheer compute requirements}.
We overcome the first challenge by avoiding producing \ac{HDR} directly.
Instead, we produce a set of individual \emph{brackets}, \ie \ac{LDR} images, which can be merged into an \ac{HDR} image.
This allows us to circumvent the first challenge by never operating the denoiser on \ac{HDR} images, and hence, also overcome the second challenge, as we circumvent the need to re-train the denoiser in \ac{HDR}.
Our method does not need any fine-tuning or training and considers the denoiser a black box.

\myfigure{Merging}{
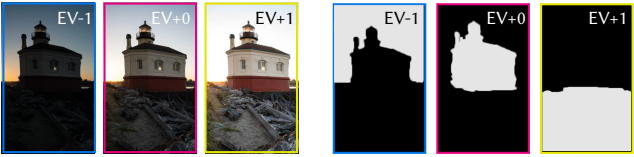
Recalling HDR merging:
LDR brackets are shown on the left;
right, the weights for each bracket, for simplicity in binary.
White means this pixel will contribute to the final HDR.}

Instead, the task is to produce brackets that are meaningful, \ie meaningful on their own and meaningful in combination with other brackets (\refFig{Merging}).
To be plausible on its own, a bracket should have all details, without noise, in the range of values it represents.
To work as a combination, a value in one bracket must match its value re-exposed to another bracket and ultimately when they are merged.
We achieve these properties by deriving a diffusion process based on ideas from \ac{DPS}~\cite{chung2022diffusion} that operates between multiple brackets jointly.

\mysection{Background: Multi-exposure HDR imaging}{Background}

HDR images directly register scene radiance, typically up to a scale factor, so that image details in the darkest and brightest scene regions are readily available.  
As sensors with HDR capabilities are relatively rare and expensive, typically, a stack of differently exposed LDR photographs (refer to \refFig{Merging}) is merged into an HDR image \cite{debevec1997recovering,Mitsunaga1999RadiometricSC,robertson2003estimation,wang2023neural}.
By transforming each pixel value through an inverted camera response and then dividing by the exposure time, a measurement of the scene radiance can be derived \cite{reinhard2010high}.
As such, per-pixel measurements are the most reliable in the middle range of the camera response  \cite{debevec1997recovering}; an accordingly weighted average of the measurements can be computed for all exposures.
\refFig{Merging}-right shows a simplified version of such weights for exposure brackets EV-1, EV+0, and EV+1, where EV+$x$ denotes multiplying with $2^x$ in the linear radiance space.
Note that the radiance ranges below the black level and over 1 are covered just in a single exposure EV+1 and EV-1, respectively, while for EV+0, radiance information is clamped on both sides of the range.
Dark image regions are also contaminated with sensor noise, whose characteristics may differ between exposures, which makes consistent denoising difficult \cite{mustaniemi2020lsd$_2$,chang2020low,cogalan2022learning}.
Some camera manufacturers introduce hard clamping at a black-level radiance, assuming that there is no reliable image information below this threshold due to noise.
Finally, the performance of the multi-exposure methods might be limited for large scene/camera motion that causes ghosting that is further aggravated by simultaneous image saturation \cite{kalantari2017deep,yan2019attentionguided,yan2020ghost,wu2018deep}.
The latter problem can be reduced through consistent image hallucination using adversarial training \cite{niu2021hdr,li2022uphdr-gan} or conditional diffusion \cite{yan2023towards} components.

In this work, we aim to use diffusion  \cite{ho2020denoising,sohl2015deep,chung2022diffusion} to generate consistent multiple exposures.
In this process, we need to account for missing information due to clamping 
and, when relevant, denoise.

\mysection{Previous Work}{PreviousWork}

In this section, we discuss previous work on deep single-image HDR reconstruction methods and the use of diffusion models in HDR imaging that are central to this work.
A broader perspective on other aspects of deep learning for HDR imaging can be found in a recent survey \cite{wang2022deep}.

\paragraph*{Deep single-image HDR reconstruction (LDR2HDR)}

An alternative solution to multi-exposure techniques (\refSec{Background}) relies on restoring HDR information from a single LDR image.
Traditional methods are extensively covered by Banterle \etal \cite{banterle2017advanced}, and here, we focus on recent machine-learning solutions.  
Single-image HDR reconstruction can be performed directly  \cite{eilertsen2017hdr, marnerides2018expandnet, santos2020single, liu2020single, zhang2021deep, yu2021luminance,chen2022text2light}, or, alternatively, by first producing a stack of different exposures that are then merged into an \ac{HDR} image~\cite{endo2017deep,lee2018deep,lee2018re,lee2020learning,jo2021deep}.
Instead of producing LDR stacks with fixed predefined EVs, Chen \etal \cite{chen2023learning} propose generating LDR stacks at continuous arbitrary values to achieve higher quality.
Specialized solutions are required when an observation EV+0 is captured in dark conditions, where denoising is a key problem \cite{chen2018learning,wang2023exposurediffusion}.
Text conditioning driven by a contrastive language-image pre-training (CLIP)  model \cite{radford2021learning} can be used for the generation of a well-exposed LDR environment map that is then transformed into its HDR counterpart by a fully supervised network \cite{chen2022text2light}.
Even though some methods employ adversarial training ~\cite{zhang2021deep,lee2018re}, the key problem remains limited performance in reconstructing clamped regions.
Those methods mostly require LDR and HDR image pairs for training, which is problematic due to limited datasets.
Recently, GlowGAN \cite{wang2023glowgan} addressed the latter two problems by fully unsupervised learning a generative model of HDR images exclusively from in-the-wild LDR images.
As this approach is based on StyleGAN-XL~\cite{sauer2022stylegan}, it requires GAN training on narrow domains (\eg lightning, fireworks) to capture the respective HDR image distribution. 
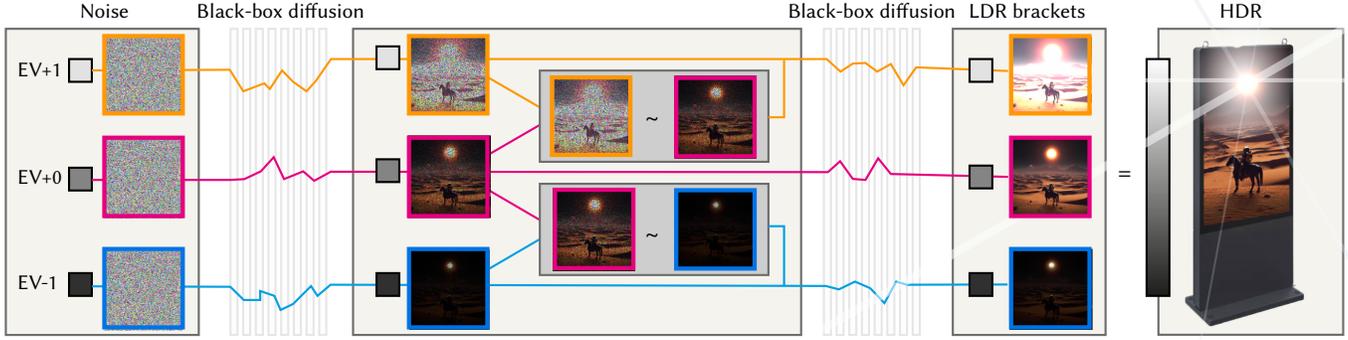
\mycfigure{Concept}{Overview of our approach.
Diffusion occurs from left to right and across multiple exposure levels (brackets), shown vertically.
We show an example with three brackets.
The process starts with three independent noises.
At each diffusion step (one is shown), denoising is guided by an brackets consistency term (middle block).
In this term, first, a denoised estimate of the current noisy images is computed (\refEq{current_estimate}), then
brackets are made consistent when re-exposed ($\sim$ symbol) using \refEq{costDown} and \refEq{costUp}.
When diffusion has finished, the brackets form an HDR image under a common HDR fusion technique.}

\paragraph*{Diffusion models in HDR imaging}
\Acp{DDPM}~\cite{ho2020denoising,sohl2015deep} demonstrate huge capacity in modeling complex distributions and typically outperform other generative models in terms of image realism, diversity, and detail reproduction \cite{dhariwal2021diffusion}.  
\acp{DDPM} also proved useful for solving linear \cite{song2021scorebased} and non-linear \cite{chung2022diffusion} inverse imaging problems that are common in image restoration and enhancement tasks guided by the degraded input image.
Image inpainting \cite{lugmayr2022repaint}, deblurring \cite{kawar2022denoising}, and super-resolution \cite{Saharia2023image} are examples of such restoration tasks, where the degradation models are typically linear and known \cite{fei2023generativediffusion}.
In HDR imaging tasks, the degradation model is more complex, and existing solutions based on \acp{DDPM} are more sparse.
Wang \etal \shortcite{wang2023exposurediffusion} propose low-light image enhancement using exposure diffusion that is directly initialized with the noisy low-light image instead of Gaussian noise, which greatly simplifies denoising and consequently reduces the network complexity and the required number of inference steps.
The method can be trained using pairs of low-light and normally-exposed photographs, as well as synthetic data using different noise models.
Fei \etal \shortcite{fei2023generativediffusion} employ a pre-trained \ac{DDPM} and propose the Generative Diffusion Prior (GDP) for unsupervised modeling of the natural image posterior distribution. 
They demonstrate the utility of this framework for low-light image enhancement and HDR image reconstruction by merging low, medium, and high exposures.
A similar task, but with explicit emphasis on large motion between the three exposures and severe clamping at the same time, is addressed in Yan \etal \cite{yan2023towards}.  
Lyu \etal \cite{lyu2023diffusion} train a \ac{DDPM} to capture the distribution of natural HDR environment maps, but are limited to rather narrow classes (\eg urban streets) due to scarcity of available HDR training data.
Dalal \etal \shortcite{dalal2023single} train a \ac{DDPM} on LDR--HDR image pairs (roughly 2,000 images, from the HDR-Real~\cite{liu2020single} and HDR-Eye~\cite{nemoto2015visual} datasets) and reconstruct HDR images from single LDR images.

Our work follows Chung \etal \cite{chung2022diffusion} and relies on off-the-shelf pre-trained diffusion models 
\cite{dhariwal2021diffusion,nichol2021glide} 
that feature better domain generalizability due to intensive training on large datasets than explicit training on small datasets of LDR--HDR image pairs \cite{dalal2023single,lyu2023diffusion}.
Our solution does not require any HDR images at the training stage.
Instead, we implicitly leverage the exposure statistics of real-world photographs used for \ac{DDPM} training, which allows the model to reason on the underlying radiance distributions.
In single-image reconstruction, we require as the input just one LDR exposure and then generate a stack of different spatially consistent LDR exposure brackets.
This way, we avoid possible problems with large motion inherent for time-sequential capturing \cite{fei2023generativediffusion,yan2023towards}.

Optionally, the hallucinated HDR content in saturated regions can be conditioned on text prompts \cite{nichol2021glide}.
Such text prompts can also be used as the only input to generate standalone HDR images. 
Histograms with the desired pixel color distribution, possibly derived from existing images, can guide global contrast relations in generated HDR content and can optionally be combined with text prompts.
\refTab{combinations} summarizes all text conditioning and image/histogram guidance combinations we explore.
With respect to non-diffusion methods such as GlowGAN \cite{wang2023glowgan}, we benefit from an overall better quality of generated images by diffusion models \cite{dhariwal2021diffusion,nichol2021glide} and avoid a lossy inversion of an input LDR exposure into a latent code as required by GANs.

Our approach also differs from existing methods that enforce consistency between multiple joint diffusion instances to create seamless high-resolution panoramas by blending colors, features \cite{bar-tal2023multidiffusion,jimenez2023mixture}, maintaining style and content \cite{lee2023syncdiffusion}, or ensuring semantic coherence \cite{quattrini2024mergingsplittingdiffusionpaths}.  In contrast, our work focuses on bracket consistency requirements specifically for HDR reconstruction.
In \refFig{high_res}, we demonstrate how HDR-specific conditions can also be combined with panorama stitching consistency.


\mysection{Our Approach}{Our Approach}

We will first briefly recall the mechanics of sample generation using \acp{DDPM} with a guiding term (\refSec{Diffusion}), before presenting our idea (\refSec{ExposureDiffusion}).

\mymath{\bracket}{\mathbf x}
\mymath{\bracketLatent}{\mathbf z}
\mymath{\allBrackets}{X}
\mymath{\positiveNumberOfBrackets}{n}
\mymath{\negativeNumberOfBrackets}{m}
\mymath{\iteration}{t}
\mymath{\denoiser}{\epsilon}
\mymath{\score}{\mathbf{s}_\theta}
\mymath{\scoreLatent}{\mathbf{s}_\theta^*}
\mymath{\decoder}{\mathbf{D}}
\mymath{\schedule}{{\alpha_\iteration}}
\mymath{\conditionsignal}{\mathbf{c}}
\mymath{\guidingsignal}{\mathbf{y}}
\mymath{\cost}{C}
\mymath{\exposure}{\alpha}
\mymath{\wiener}{\mathbf{z}_\iteration}
\mymath{\consistency}{\mathtt{braco}}
\mymath{\posteriorBracket}{\hat\bracket}
\mymath{\posteriorBracketLatent}{\hat\bracketLatent}
\mymath{\allPosteriorBracket}{\hat\allBrackets}
\mymath{\oneBracket}{\posteriorBracket^i}
\mymath{\referenceBracket}{\posteriorBracket^{r}}
\mymath{\costUp}{C_\uparrow}
\mymath{\CRF}{CRF_{\gamma}}
\mymath{\invCRF}{CRF^{-1}_{\gamma}}
\mymath{\costDown}{C_\downarrow}
\mymath{\costMain}{C_0}
\mymath{\mainFunction}f

\definecolor{satcolor}{rgb}{0.2, 0.6, 0.9} 
\definecolor{darkcolor}{rgb}{0.9, 0.2, 0.1} 
\mymath{\sat}{\mathtt{\color{satcolor}{sat}}}
\mymath{\dark}{\mathtt{\color{darkcolor}{dark}}}

\mysubsection{Guided Diffusion}{Diffusion}
Data generation with a pre-trained \ac{DDPM}~\cite{ho2020denoising,sohl2015deep} amounts to gradual denoising of a sample $\bracket \in \mathds{R}^u$ using
\begin{equation}
    \label{eq:ode}
    \bracket_{\iteration-1}
    :=
    \frac{1}{\sqrt{\schedule}}
    \Bigl(
        \bracket_{\iteration}
        -
        (1 - \schedule)
        \nabla_{\bracket_\iteration} \log p_\iteration(\bracket_\iteration)
    \Bigl)
    +
    \wiener.
\end{equation}
This update rule involves a noise schedule $\schedule \in \mathds{R}_+$, random vectors $\wiener \in \mathds{R}^u$, and, at its core, a score function 
$\nabla_{\bracket_\iteration} \log p_\iteration(\bracket_\iteration)$.
Optionally, the score can be conditioned on a signal $\conditionsignal \in \mathds{R}^v$, such as a text prompt embedding, to yield  
$\nabla_{\bracket_\iteration} \log p_\iteration(\bracket_\iteration | \conditionsignal)$.
In modern \acp{DDPM}, scores are typically approximated by a neural network 
$\score(\bracket_\iteration, \conditionsignal, \iteration) \in (\mathds{R}^u \times \mathds{R}^v \times \mathds{Z} ) \rightarrow \mathds{R}^u$.
Please refer to Yang \etal \cite{yang2023diffusion} for an in-depth treatise.

In the framework of \acf{DPS}~\cite{chung2022diffusion}, an additional guiding signal $\guidingsignal \in \mathds{R}^w$, such as a partial observation of \bracket, is incorporated into the denoising process to arrive at the posterior score
\begin{equation}
    \label{eq:posterior_estimate}
    \nabla_{\bracket_\iteration} \log p_\iteration(\bracket_\iteration | \conditionsignal, \guidingsignal)
    \approx
    \score(\bracket_\iteration, \conditionsignal, \iteration)
    -
    \lambda 
    \nabla_{\bracket_\iteration} \cost(\posteriorBracket_\iteration, \guidingsignal).
\end{equation}
Here, 
$\cost \in (\mathds{R}^u \times \mathds{R}^w) \rightarrow \mathds{R}$
is a problem-specific measurement term that drives the denoising process towards solutions that incorporate the guiding signal \guidingsignal, and $\lambda \in \mathds{R}_+$ is a balancing term.
For increased stability, Chung \etal \cite{chung2022diffusion} propose to feed the current estimate of the clean sample
\begin{equation}
    \label{eq:current_estimate}
    \posteriorBracket_\iteration
    =
    \frac{1}{\sqrt{\bar{\schedule}}}
    \Bigl(
        \bracket_{\iteration}
        +
        (1 - \bar{\schedule})
        \score(\bracket_\iteration, \conditionsignal, \iteration)
    \Bigl)
\end{equation}
to \cost, where $\bar\schedule$ is derived from \schedule.

\mysubsection{Exposure diffusion}{ExposureDiffusion}
The above equations \refEq{ode} and \refEq{posterior_estimate} are valid for producing a single LDR result image \bracket.
Our idea is to produce HDR by diffusing multiple LDR results.
Hence, we operate (\refFig{Concept}) on a set of LDR images
$
\{
\bracket^{-\negativeNumberOfBrackets},
\ldots,
\bracket^0,
\ldots,
\bracket^\positiveNumberOfBrackets
\}
$,
called ``brackets''.
Positive and negative superscripts denote positive and negative EVs, respectively.
All brackets are initialized to noise with mean zero and standard deviation one.
They, further, need to be gamma-corrected sRGB \ac{LDR} images, as we consider the score function a black box that cannot be retrained to work on linear \ac{HDR}.

\paragraph*{Score term}
The first term in \refEq{posterior_estimate} is the common score function that points from the current solution into the direction of a more plausible one.
It may or may not be conditioned as per the second column of \refTab{combinations}, leading to different application scenarios.
It is a black box we do not need to know any details of, nor differentiate, as it already encodes a gradient.
We only need to know its noise schedule \schedule to also use \posteriorBracket from \refEq{current_estimate}.
The score function is hence simply computed on each bracket independently.

\paragraph*{Posterior term}
The second term in \refEq{posterior_estimate} is very specific to our problem, the bracket consistency term.
The consistency of two brackets measures how much \oneBracket, a free variable, is compatible with another bracket \referenceBracket that is assumed fixed.
For each bracket \oneBracket, the reference bracket \referenceBracket is exposed to another bracket (that can both be higher or lower EV), and the resulting differences are checked using the function \consistency, defined as
\[
\consistency(
\referenceBracket
\rightarrow 
\oneBracket
) :=
\CRF\left(\min(\frac{\exposure^i} {\exposure^\mathrm r}
\odot
\invCRF(\referenceBracket), 1)\right)
-
\oneBracket
,
\]
where $ \CRF(x) = x^{\gamma}$ with $\gamma = \frac{1}{2.2}$ represents the camera response function, and its inverse is given by $ \invCRF(x) = x^{1/\gamma} $. 
We first apply inverse CRF, as the solution exists in non-linear space for the black box score. Next, we scale by the ratio between the exposure times (\exposure) and then clamp and apply CRF again to simulate the behavior of a real camera.

Since negative EVs primarily involve hallucinating saturated content and positive EVs focus on denoising, our posterior term behaves slightly differently for positive, negative, and zero EV brackets.
The posterior for decreasing exposure (negative EVs) is
\begin{gather}
\begin{aligned}
\label{eq:costDown}
\costDown(\oneBracket, \referenceBracket) 
=
& 
||
\sat(\referenceBracket)
\cdot
\max(
\consistency(\referenceBracket\rightarrow \oneBracket),
0
)
||_2
+
\\
\lambda_\mathrm s \cdot
&||
(1-\sat(\referenceBracket))
\cdot
(
\consistency(\referenceBracket\rightarrow \oneBracket)
)
||_2,
\end{aligned}
\end{gather}
while the one to increase exposure (positive EVs) is
\begin{gather}
\begin{aligned}
\label{eq:costUp}
\costUp(\oneBracket, \referenceBracket) 
=
&||
\dark(\referenceBracket)
\cdot
(
\consistency(\referenceBracket\rightarrow \oneBracket)
)
||_2
+
\\
\lambda_\mathrm d \cdot
&|| 
(1-\dark(\referenceBracket))
\cdot
(
\consistency(\referenceBracket\rightarrow \oneBracket)
)
||_2
,
\end{aligned}
\end{gather}
where $\lambda_\mathrm s $ and $\lambda_\mathrm d$ are the balancing weights. The \sat and \dark are the mask functions for saturated and near-zero pixels, respectively, and zero otherwise. However, in practice, we use linear functions $\sat(\bracket)=\bracket$ and $\dark(\bracket)=1-\bracket$ instead of conventional binary masking \cite{kalantari2017deep} to make our cost functions smooth and tractable.
The possible combinations of consistency and up or down direction are discussed with an example in \refFig{cost}.

The $\max$ operation in {\refEq{costDown}} is responsible for generating plausible content in saturated areas. To clarify its role, consider \oneBracket as the optimized EV-1 bracket for \referenceBracket. In regions where \referenceBracket is saturated (e.g., the blue dots in the top row of \refFig{cost}), there is a feasible range of values that \oneBracket can take, such that when exposed to \referenceBracket, they are clamped to 1. For the EV-1 case, this range is from 0.5 to 1. This constraint is enforced by the term $\sat(\referenceBracket) \cdot \max(\referenceBracket / 2 - \oneBracket, 0)$ \change{(assuming an identity CRF in this didactical example)}. The max term encourages the optimized bracket \oneBracket to be any value above $\referenceBracket /2$. Consequently, $\referenceBracket /2 - \oneBracket$ becomes negative, resulting in a zero cost.

The weighting factor $\lambda_\mathrm{s}$ in \refEq{costDown} is set to 1; however, in \refEq{costUp}, we weigh the two terms differently, with $\lambda_\mathrm{d} = 2$, to account for the noise removal effect. The darker regions  (\eg the red dots in the bottom row of \refFig{cost}) are often noisy or less reliable, so we apply a smaller coefficient to impose less data term prior in these areas compared to brighter regions (i.e., $1.0 - \dark(\referenceBracket)$). 
\myfigure{cost}{Posterior based on bracket consistency cost for optimizing lower exposure (top row) and higher exposure (bottom row).
The horizontal axis in the cost plot represents the pixel values in the current solution \oneBracket, and dots are placed where their value in the reference \referenceBracket is.  The vertical axis shows the cost values, with horizontal lines representing zero cost. Depending on the exposure direction, this results in different costs for choices in \oneBracket. When going down in exposure (top row), for the saturated region, we allow \oneBracket to take any value within a feasible range, such that when exposed to \referenceBracket, they will be clamped to 1. For higher exposure (bottom row), the consistency term is relaxed (indicated by a lower steepness of the penalty cost) for dark areas compared to other regions. }

Finally, we can also define an optional posterior term on the original image by applying a function \mainFunction:
\begin{gather}
\begin{aligned} 
\label{eq:cost0}
\costMain(\oneBracket, \guidingsignal)=
\lambda_c\cdot||
\mainFunction(\oneBracket)-\guidingsignal
||_2.
\end{aligned}
\end{gather}
First, if \mainFunction is, for example, the identity, and \guidingsignal an LDR image (the third column in \refTab{combinations}), this becomes a reconstruction task.
In that case, the solution for \oneBracket is immediately set to \guidingsignal. 
As a second alternative, we explore using conversion to an LDR histogram as \mainFunction. In this case, the parameter $\lambda_c$ is set to 10. 

Combining all together, we arrive at our final cost \cost:
\begin{gather}
\begin{aligned}
\label{eq:our_posterior}
\cost(\oneBracket,\guidingsignal)
=
\begin{cases}
\costDown(\oneBracket,\posteriorBracket^{i+1})
&, \text{if }i<0 
\text{, see \refEq{costDown},}\\
\costUp(\oneBracket,\posteriorBracket^{i-1})
&, \text{if }i>0
\text{, see \refEq{costUp} and}\\
\costMain(\oneBracket, \guidingsignal)
&, \text{if }i=0
\text{, see \refEq{cost0}.}
\end{cases}
\end{aligned}
\end{gather}

\refEq{our_posterior} is the expression for a single exposure bracket \oneBracket.
As per \refEq{posterior_estimate}, this expression gets differentiated with respect to its first argument.
The subtlety is that this is now done for multiple brackets, but they depend on each other.
In our implementation, during one optimization step, however, for each bracket, the other bracket \referenceBracket is considered a constant, so the second argument of \costDown, \costUp, and \costMain is ``detached'' in PyTorch parlance.
Note that this is different from greedily optimizing each bracket sequentially.

\begin{table}[]
    \centering
    \setlength{\tabcolsep}{2.5pt}
    \caption{\change{Our method supports various applications through different combinations of score conditioning (text or null) and guidance (image, histogram, or none). For reconstruction tasks, the EV+0 is fixed to the input LDR image. The final column specifies the diffusion backbone used. Please note our approach is model-agnostic, meaning it can be adapted to different diffusion models based on the application. For instance, we utilize GLIDE's conditional model \cite{nichol2021glide} for text-conditioned experiments and Stable Diffusion \cite{rombach2022high-resolution} for generating high-resolution samples.}}
    \label{tab:combinations}
     \resizebox{\columnwidth}{!}{%
    \begin{tabular}{lccccc}
    \toprule
        \multicolumn1l{Application}&
        \multicolumn1c{Cond. \conditionsignal   }&
        \multicolumn1c{Guide \guidingsignal}&
        \multicolumn1c{EV+0 fix?}&
        \multicolumn1c{Example}&
        \multicolumn1c{\change{Backbone model}}

        \\
        \midrule
        Generation&
        Text&
        ---&
        \xmark&
        \refFig{text2HDR}, \ref{fig:text2HDRLatent}&\cite{nichol2021glide,rombach2022high-resolution}\\
        Generation&
        ---&
        Histo.&
        \xmark&
        \refFig{histogram}& \cite{nichol2021glide}\\
        Generation&
        Text&
        Histo.&
        \xmark&
        \refFig{text_Histogram}&\cite{nichol2021glide}\\
        \midrule
        Recons.&
        ---&
        Image&
        \cmark&
        \refFig{constant_lambda}, \ref{fig:LDR2HDR}, \ref{fig:denoising}, \ref{fig:high_res} & \cite{dhariwal2021diffusion}\\
        Recons.&
        Text&
        Image&
        \cmark&
        \refFig{text_ldr}&\cite{nichol2021glide}\\
        \bottomrule
    \end{tabular}}
    \vspace{-.3cm}%

\end{table}

\mysection{Results}{Results}

\newcommand{\method}[1]{\texttt{\textbf{#1}}}

We begin by describing our experimental setup in \refSec{setup}. We then showcase the application of our method to HDR generation (\refSec{Generation}) and reconstruction (\refSec{Reconstruction}), providing quantitative as well as qualitative results for both tasks.

\mysubsection{Experimental setup}{setup}
For our reconstruction experiments, specifically the LDR2HDR task, we utilize the pre-trained image-domain unconditional diffusion model of Dhariwal \etal \cite{dhariwal2021diffusion}. Our input images are down-sampled to 256$\times$256 before they are fed to this model, and we perform T=1,000 denoising steps to produce our results.
In tasks involving text-conditioning or histogram guidance, we use the OpenAI GLIDE \cite{nichol2021glide} diffusion model, which is text-conditional and generates images at a resolution of 64×64 using a classifier-free guidance strategy. Subsequently, an upsampling diffusion model is applied to increase the resolution to 256×256. In this case, we apply our DPS approach only to the text-conditional model and perform T=500 steps to produce the results. Once the exposure brackets are generated, they are individually upsampled using \change{GLIDE's pre-trained} upsampling module.

The hyper-parameter $\lambda$ in {\refEq{posterior_estimate}} balances between the diffusion prior and our posterior term. 
\change{It is worth noting that saturated regions are also included in our posterior term {(\refEq{costDown})}, and since $\lambda$ determines the weight of this term, its value directly affects the hallucinated content.} We set $\lambda=1.5$ when employing the conditional diffusion model \cite{nichol2021glide}.
However, in our experiments with the unconditional diffusion model \cite{dhariwal2021diffusion}, we observe that a constant $\lambda$ sometimes leads to unrealistic hallucinations for saturated regions, as shown in \refFig{constant_lambda}. To achieve more consistent hallucinations, we adopt a time-dependent weight $\lambda = \lambda_0 \cdot (1 - t/T)^2$ with $\lambda_0 = 6$. Intuitively, each bracket is initialized randomly at the beginning, making it difficult for the data consistency term to provide the correct gradient. Therefore, we reduce its influence at the beginning ($t = T$) and gradually increase it as the denoising progresses. 


For all results, we compute five exposure brackets: EV-4, EV-2, EV+0, EV+2, and EV+4, unless otherwise specified.
These exposure brackets are merged using the standard technique \cite{debevec1997recovering} to create our HDR image.
For \change{\refFig{LDR2HDR}, \ref{fig:text_ldr}, and \ref{fig:denoising}}, we show the result by applying the tonemapping of Mantiuk \etal \cite{mantiuk2006perceptual} while in all other results, we directly show the optimized brackets. \change{We release our code and provide the results in an HDR format on our webpage: https://bracketdiffusion.mpi-inf.mpg.de/ }

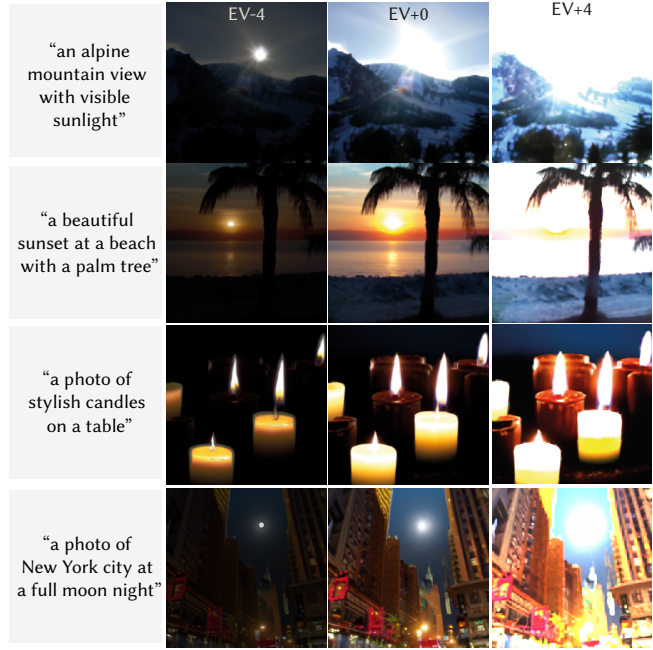
\myfigure{text2HDR}{Text-based HDR generation.
Text prompts are on the left, alongside low (EV-4), medium (EV+0), and high exposures (EV+4).}

\mysubsection{Generation}{Generation}
Image generation is a premiere ability of diffusion models, which we extend to HDR.
Image generation without any conditioning or guidance frequently results in scenes that, in reality, do not exhibit high dynamic ranges. 
Therefore, capitalizing on the generality of our framework, we consider generation conditioned on text prompts, guided by RGB color histograms, and a combination thereof (first three rows in \refTab{combinations}).

\paragraph*{Text-based}
Here, we consider the task of text-conditioned generation, where the score function takes a conditioning signal \conditionsignal in the form of a text embedding.
We omit \costMain, \ie the generation is free to synthesize any consistent brackets following the text prompt.
Results of this application are shown in \refFig{text2HDR}.
The low exposures present detailed depictions of visible light sources, such as the structure of candle flames, including glares typically found around strong light sources.
In the daylight scenes, most of the details are properly exposed for the medium exposure (EV+0), while in the night scenes, a high exposure (EV+4) is required to see sufficient detail.
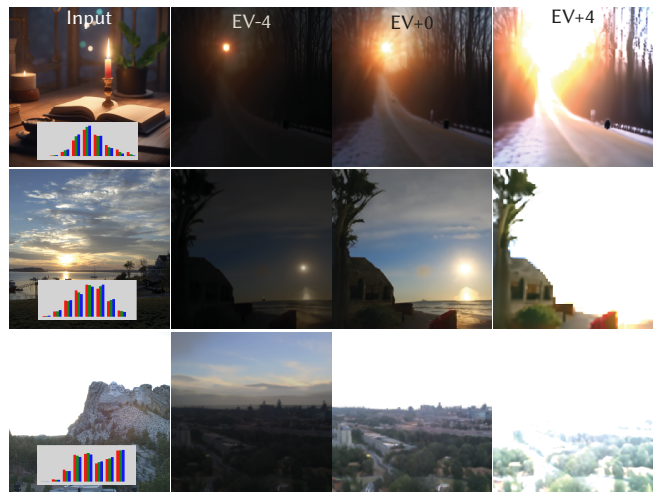
\myfigure{histogram}{Histogram-based HDR generation.
The first column shows the input image and its histogram.
The other columns show our generated brackets.
Note that the method never sees the input image (left), only its histogram.}
\paragraph*{Histogram-based}
Here, we explore guided generation using a target histogram.
In our experiments, we first compute an LDR histogram with 10 bins per color channel of an input image as our guiding signal \guidingsignal (\refFig{histogram}, first column). 
Then, we utilize \costMain to direct the generation process towards producing an EV+0 bracket that matches this histogram (\refFig{histogram}, third column), using a differentiable histogram function with soft bin assignments as \mainFunction.
Our framework produces consistent brackets of HDR content (\refFig{histogram}, second to fourth column).

\paragraph*{Text \& histogram-based}
In \refFig{text_Histogram}, we combine the control modalities of the previous two paragraphs.
In the first three rows, we apply constraints where 50\%, 25\%, and 1\% of saturated pixels are enforced on the histograms of the EV+0 bracket, all while utilizing the same text prompt.
We observe that our approach enables the generation of different HDR contents that faithfully reflect the queries.
In the last row, a guiding histogram is extracted from an input image.

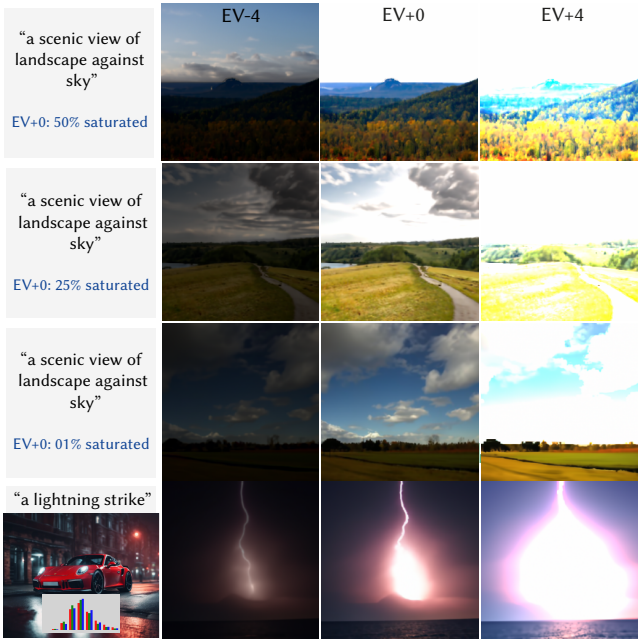
\myfigure{text_Histogram}{Text- and histogram-based HDR generation.
The first column is the query, and the other three columns are our results.  Additional results are provided in our supplementary. }

\myfigure{constant_lambda}{The effect of different $\lambda$ setting in \refEq{posterior_estimate} on the LDR2HDR task. The reconstructed (tone-mapped) HDR results are shown on the right for a given input LDR image (left). A constant $\lambda$ value often leads to reconstructions with artifacts, whereas our proposed time-dependent setting, $\lambda = \lambda_0 \cdot (1- t/T)^2$ (See \refSec{setup}), produces significantly better results.}

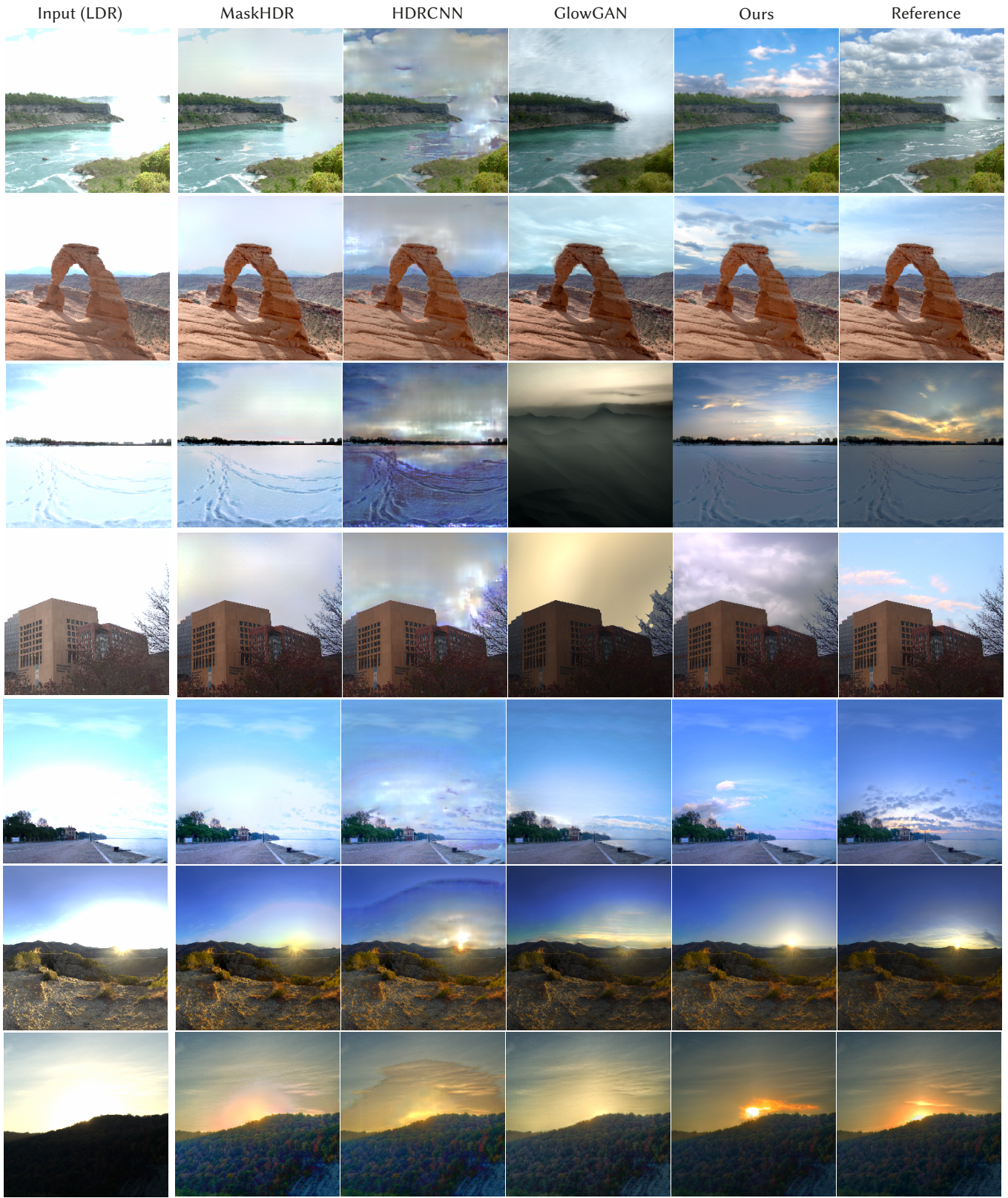
\mycfigure{LDR2HDR}{LDR2HDR reconstruction for our method and competitors given an input LDR images (first column). All HDR images (right columns) are tone-mapped using the same tone-mapper, whose parameters are tuned for each row to achieve the best visual appearance of the corresponding reference HDR image.}

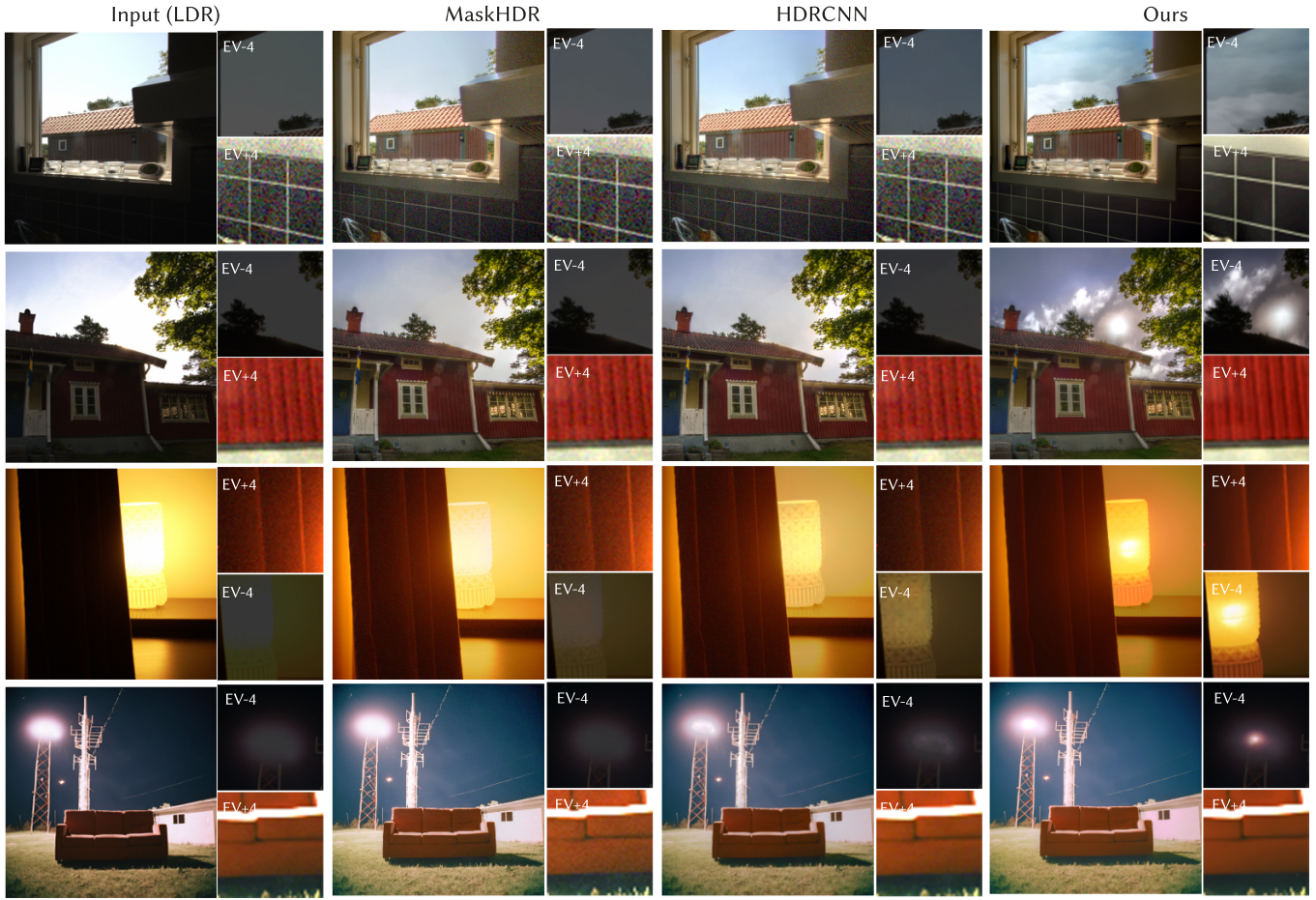
\mycfigure{denoising}{LDR2HDR reconstruction for MaskHDR, HDRCNN, and Ours methods guided by the input LDR images (left column).
Insets show dark, and hence noisy, as well as bright, partially saturated input regions.
Other methods can remove some noise, but ours not only gets the semantics right in saturated areas (\eg for the lamp or sun), but also removes noise in dark areas. 
The images in the first three rows are examples from the SI-HDR dataset \cite{hanji2022comparison}, while the input image in the last row is an AI-generated image with Stable-Diffusion.}

\mysubsection{Reconstruction}{Reconstruction}
We now turn to one of the supreme disciplines of HDR imaging: LDR2HDR restoration.
There are two major challenges involved in this task.
Firstly, we need to fill the saturated (white) regions in the LDR image \guidingsignal with appropriate content. 
Secondly, dark regions in \guidingsignal often contain strong noise that needs to be removed.
Our approach naturally supports this task by setting \mainFunction in \refEq{cost0} to be the identity function.
We demonstrate both unconditional and text-conditioned reconstruction (last two rows in \refTab{combinations}).

\paragraph*{Methods and dataset}
We compare our approach for the LDR2HDR task with \method{CERV} \cite{chen2023learning} and \method{GlowGAN} \cite{wang2023glowgan}, which are recent state-of-the-art methods. Additionally, we evaluate against two other top-performing methods, \method{MaskHDR} \cite{santos2020single} and \method{HDRCNN} \cite{eilertsen2017hdr}, as identified in recent studies \cite{banterle2024self, wang2023glowgan}. 
Note that the only other generative approach, \method{GlowGAN}, requires training a domain-specific model. Thus, for a fair comparison, we limit our evaluation to landscape images, as a pre-trained \method{GlowGAN} model is available for this category.
Specifically, we curate a dataset comprising 75 HDR images sourced from various online platforms, which will be made available on publication.

\paragraph*{Metrics}
We employ \change{four} different metrics to assess restoration performance.
Firstly, we employ the full-reference metric HDR-VDP-3~\cite{mantiuk2023hdr}, which evaluates reconstruction fidelity without considering that saturated regions in an LDR image may allow for multiple, different HDR solutions.
Secondly, to gauge overall plausibility, we utilize the no-reference HDR image metric PU21-PIQE~\cite{hanji2022comparison}.
This metric, however, is agnostic of the expected distribution of hallucinated contents in our narrow domain. 

\change{To address these considerations, we also employ two additional metrics: DreamSim \cite{fu2023dreamsim} and FID \cite{heusel2017gans}. DreamSim evaluates high-level visual similarities and differences between image pairs, providing insights into perceptual alignment. Meanwhile, the FID score, widely used in generative settings, measures discrepancies between distributions of generated and reference images, serving as a reliable measure of generative quality.}
However, since FID relies on a vision model \cite{krizhevsky2012imagenet} pre-trained on LDR images, it cannot be directly applied to HDR content.
Rather, we seek to produce a representative distribution of LDR images derived from the HDR content, accounting for uncalibrated and unnormalized pixel values across methods.
We opt to apply the auto-exposure method by Shim \etal \cite{shim2014auto} to each HDR image. 
This technique helps determine the EV0 bracket, from which we derive EV$\pm$2 and EV$\pm$4 brackets. 
Subsequently, we select 100 random 64$\times$64-pixel crops from each image. We maintain consistency in selecting crop locations across methods \cite{chai2022any}. 
This precaution is necessary because having small bright light sources, such as the sun, in some patches in one method but not in another could disproportionately bias the measurement.
Our protocol leads to stable estimates based on 7.5k patches per bracket and 37.5k patches in total.

\newcommand{\query}[1]{\texttt{#1}}
\begin{table}[]
    \setlength{\tabcolsep}{2pt}
    \caption{Reconstruction task performance. The first and second best-performing methods are highlighted in bold and underlined, respectively. $\method{Ours} ^{\dag}$ refers to a version of our method with a more complex camera response function (see \refSec{Ablations}).}
    \label{tab:Generation}
    \centering    
    \resizebox{\columnwidth}{!}{%
    \begin{tabular}{lc ccccccccc}
    \toprule
    &
    \multicolumn6c{FID$\downarrow$ }&
 &&
    \\
    \cmidrule(lr){2-7}
    \multicolumn1l{Method}&
    \multicolumn1c{EV-4}&
    \multicolumn1c{EV-2}&
   \multicolumn1c{EV+0}&
   \multicolumn1c{EV+2}&
   \multicolumn1c{EV+4}&
    \multicolumn1c{All.}&
   \multicolumn1c{\change{DreamSim$\downarrow$}}&
    \multicolumn1c{No-Ref.$\downarrow$}&
    \multicolumn1c{Full-Ref.$\uparrow$}
    
     \\
    \midrule
    \method{MaskHDR} &  14.36 & 09.44 & \textbf{04.13} & \textbf{01.14} & \textbf{02.81} & 03.63 & \underline{0.053}&51.7 $\pm$ 7.5 &  05.87 $\pm$ 1.6\\  
    \method{HDRCNN} &  14.54  & 16.89 & 13.06& 03.73 & 03.27 & 06.54& 0.082 &\underline{47.2 $\pm$ 7.1} & \textbf{06.67 $\pm$ 1.2}\\ 
    \method{CERV} & 21.83 &16.63&10.04&08.29&16.22&08.00 & 0.129 & 75.1 $\pm$  9.6& 05.14  $\pm$ 1.5\\
    \method{GlowGAN} & \underline{ 08.59}  & \underline{06.94} &05.32 & 03.61 & 08.09 & \underline{03.08} & 0.078 &\textbf{45.5 $\pm$ 8.6} & \underline{06.57 $\pm$ 1.5}\\
    \midrule
    $\method{Ours} ^{\dag}$ & 10.13 & 09.45 & 06.43 & 03.23 & 06.63 & 03.41 & 0.081 & 50.8 $\pm$ 8.1 &  06.46 $\pm$ 1.3\\ 
    \method{Ours} & \textbf{ 06.25} & \textbf{06.48}& \underline{04.65} & \underline{01.28} & \underline{02.89} &\textbf{02.05} & \textbf{0.048} & 51.7 $\pm$ 7.6 & 06.51 $\pm$ 1.2\\ 
    \bottomrule
    \end{tabular}}
     \vspace{-.5cm}
\end{table}

\paragraph*{Results}
Our quantitative evaluation results are presented in \refTab{Generation}.
We observe that our approach outperforms the baselines in terms of overall FID (denoted as "All") and excels in the challenging cases of negative EV where content needs to be hallucinated.
\change{Additionally, our method achieves the best performance across all baselines when evaluated using the DreamSim metric.} Results for the other two metrics remain inconclusive due to statistical insignificance.
Note that the full-reference metrics (included here only to follow the previous practice) favor blurriness in hallucinated content and poorly evaluate its naturalness. FID, a standard metric for generative methods, clearly shows that our solution consistently outperforms all other approaches.

In \refFig{LDR2HDR}, we show corresponding qualitative results with a focus on saturated regions; complete sets of images are provided in the supplemental.
Our approach consistently generates arguably the highest-quality hallucinations in saturated regions.
This is facilitated by the first term in \refEq{costDown}, which gives the process the freedom to generate any content as long as it is bright enough.
Notably, in the third row of \refFig{LDR2HDR}, we present a particularly challenging case where one color channel is nearly entirely saturated across the image. In this instance, we observe how the baselines struggle to produce plausible content, even \method{GlowGAN}, which typically excels in generating realistic results due to its domain-specific generative capabilities.
In the last two rows, we see that \method{HDRCNN} and \method{MaskCNN} struggle with image regions close to the sun, producing unnatural discontinuities and halo effects, respectively.
\method{CERV} fails in almost all examples, which is not surprising given that the authors explicitly noted their method's inability to generate reasonable content in largely saturated regions. As anticipated, given the inherent ambiguities of the LDR2HDR restoration task, all methods, including ours, generate results that diverge from the reference.


Another challenging aspect of LDR2HDR reconstruction involves eliminating noise from regions that were initially very dark.
A na\"ive scaling of the original image content leads to substantial noise, making these results practically unusable.
In \refFig{denoising}, we illustrate how our approach serves as an effective denoiser, yielding visually pleasing outcomes.

\begin{table}[]
    \centering
    \setlength{\tabcolsep}{2.5pt}
    \caption{\change{Performance comparison in terms of runtime and GPU memory usage using a single NVIDIA Quadro RTX 8000 GPU for a 256×256 resolution input.}}
    \label{tab:timing}
    \resizebox{0.60\columnwidth}{!}{%

    \begin{tabular}{lrc}
    \toprule
        Method&Runtime&Memory(GB)
        \\
        \midrule
        \method{HDRCNN} & 0.03\,s & 2.5\\
        \method{MaskHDR}& 0.53\,s &0.5  \\
        \method{CERV}& 0.32\,s & 0.2\\
        \method{GlowGAN}  & 15\phantom{.}min & 8.0\\
        \method{Ours} w/\cite{dhariwal2021diffusion} & 22\phantom{.}min & 23.0 \phantom{ } \\
        \method{Ours} w/\cite{nichol2021glide} & 2\phantom{.}min & 9.3 \\

        \bottomrule
    \end{tabular}}
    \vspace{-.5cm}%
\end{table}

\change{We also evaluate the runtime and GPU memory usage of our method against other baselines on a single NVIDIA Quadro RTX 8000 GPU for a 256×256 resolution input, with results presented in {\refTab{timing}}. The reported runtime for our method is based on generating five brackets. As expected, diffusion-based models are significantly slower than feed-forward methods. However, using modern GPUs like the NVIDIA Tesla A100 reduces the runtime for generating five brackets with Dhariwal \etal \cite{dhariwal2021diffusion} model to approximately six minutes. Our approach also scales linearly with the number of brackets in terms of GPU memory usage. For example, using the GLIDE model \cite{nichol2021glide}, generating 3, 5, 7, and 9 brackets requires approximately 6.2, 9.3, 12.7, and 15.3\,GB of GPU memory, respectively.}

\paragraph*{Text-based reconstruction}
Our framework offers a unique opportunity: the ability to dictate which content to hallucinate in saturated regions through text conditioning.
This is demonstrated in \refFig{text_ldr}, where, in addition to the guiding LDR signal \guidingsignal, the user provides a text prompt conditioning signal \conditionsignal.
We see that this combination of control modalities enables precise HDR content generation.
We emphasize that this task differs from typical inpainting in the LDR domain. Here, saturated pixel values are not replaced by darker ones but rather extended in dynamic range while forced to align with the LDR observation (\refEq{costDown}).

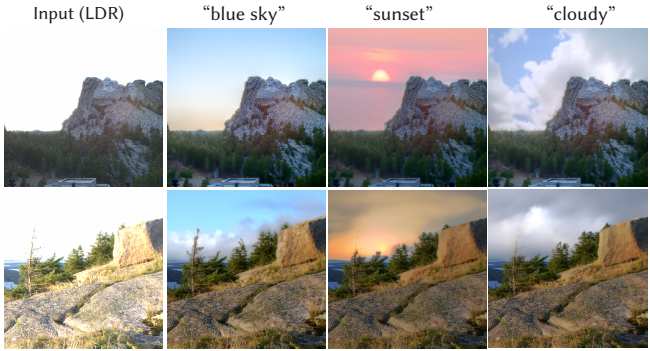
\myfigure{text_ldr}{Text-based reconstruction.
The LDR image on the left has ambiguous regions, \eg the sky.
The right three columns show what the sky could look like in a tone-mapped result on a reconstructed HDR. Each variant is conditioned on different text prompts shown on the top.}

\mysection{Ablations}{Ablations}
In this section, we analyze various aspects of our method, including \change{the number of optimized brackets}, the effect of the CRF model, the underlying pre-trained diffusion model, and different optimization strategies.

\paragraph*{Number of brackets}
\change{Our method is flexible with respect to the number of exposure brackets. We conduct two experiments to assess the impact of different numbers of brackets on output quality for the LDR2HDR task. In the first, we fix the dynamic range and vary the number of brackets, corresponding to different levels of overlap between exposures. In the second, we increase dynamic ranges while keeping the exposure ratio fixed. For both, we report FID and DreamSim scores. Additionally, to evaluate the effectiveness of our bracket consistency term, we compute the consistency between neighboring brackets by re-exposing all synthesized brackets to their neighboring ones using a process similar to our $\consistency$ function and measuring the differences using the PSNR metric.

In the first experiment, we fix the exposure range from EV-4 to EV+4 and use 3, 5, and 7 brackets. The results are summarized in {\refTab{ablation2}}. Here, the FID score is measured using the same evaluation set as in {\refTab{Generation}}. With only three exposures (EV-4, EV+0, EV+4), the optimization becomes more challenging due to inadequate sampling of the dynamic range. The best performance is achieved with five brackets, yielding the lowest FID (2.05) and DreamSim (0.048) scores, along with a bracket consistency of 39.4 dB. This level of consistency is comparable to the differences observed in high-quality JPEG compression, which is commonly used for HDR bracket fusion.
However, increasing the number of brackets to seven does not improve HDR recovery.
Our bracket consistency remains high overall; however, as the brackets are optimized recursively, with more brackets, consistency begins to decrease.

In the second experiment, we optimize for different dynamic ranges—EV-2 to EV+2, EV-4 to EV+4, and EV-6 to EV+6—with 3, 5, and 7 brackets and an EV-2 stop separation, respectively. In this experiment, we choose a subset of our evaluation set featuring an extremely high dynamic range (e.g., the presence of the sun). 
We report both per-exposure and overall FID scores in {\refTab{ablation1}}. We limit the results to exposures up to EV+2, as the outputs at EV+4 and EV+6 are nearly saturated. Overall, the findings indicate that increasing the number of brackets consistently enhances the recovery of higher dynamic ranges (e.g., EV-6). However, five brackets strike the best balance between computational efficiency and output quality, making it the practical choice for our method.}

\begin{table}[]
    \setlength{\tabcolsep}{3.5pt}
    \caption{\change{Ablation study on the number of brackets used for LDR2HDR task. Here, we fix the exposure range and increase the overlap between the exposures. The final column reports the consistency between brackets using the PSNR metric. }}
    \label{tab:ablation2}
    \centering    
    \resizebox{0.65\columnwidth}{!}{%
    \begin{tabular}{cccc}
    \toprule
    \multicolumn1l{\#EVs}&
    \multicolumn1c{FID$\downarrow$ (All.)}&
    \multicolumn1c{DreamSim$\downarrow$}&
    \multicolumn1c{Consist.$\uparrow$ (dB)}\\
    \midrule
    3 & 03.09 & 0.063 & 39.1  \\  
    5 & 02.05 & 0.048 & 39.4  \\
    7 & 03.36 & 0.055 & 37.4 \\
    \bottomrule
    \end{tabular} }
    \vspace{-.4cm}%
\end{table}

\begin{table}[]
    \setlength{\tabcolsep}{2.5pt}
    \caption{\change{Ablation study on the number of brackets used for LDR2HDR task. Here, we extend the dynamic ranges. Bracket consistency is measured in dB.}}
    \label{tab:ablation1}
    \centering    
    \resizebox{0.90\columnwidth}{!}{%
    \begin{tabular}{ccccccccc}
    \toprule
    & \multicolumn6c{FID$\downarrow$ } \\
    \cmidrule(lr){2-7}
    \multicolumn1l{\#EVs}&
    \multicolumn1c{EV-6}&
    \multicolumn1c{EV-4}&
    \multicolumn1c{EV-2}&
   \multicolumn1c{EV+0}&
   \multicolumn1c{EV+2}&
    \multicolumn1c{All.}
    & DreamSim$\downarrow$ & Consist.$\uparrow$\\
    \midrule
    3 & 05.71 & 05.31 & 04.12 & 03.40 & 02.57 & 02.06 & 0.025 & 42.8\\  
    5 & 05.12 & 04.71 & 04.01 & 03.45 & 03.18 & 01.86 & 0.026& 38.0 \\ 
    7 &  04.48 &04.40 & 03.71 & 02.88 & 03.32 & 01.62 &0.025 & 33.4 \\
    \bottomrule
    \end{tabular} }
    \vspace{-.4cm}%
\end{table}

\paragraph*{The effect of CRF}
The CRF maps raw sensor readings, which correspond to actual light intensity, to pixel values in the displayed image. In our experiments, we employ a commonly used CRF modeled as a simple gamma function, $\CRF(x) = x^{\gamma}$. \change{Substituting this gamma function into the $\consistency$ 
consistency expression {(\refSec{ExposureDiffusion})} yields:}
\[
\consistency(\referenceBracket\rightarrow \oneBracket) :=
\left(\min(\frac{\exposure^i} {\exposure^\mathrm r}
\odot (\referenceBracket) ^ {1 / \gamma}, 1)\right)^{\gamma}
-\oneBracket.
\]
\change{ This expression can be further simplified to:}
\[
\consistency(\referenceBracket\rightarrow \oneBracket) :=
\min((\frac{\exposure^i} {\exposure^\mathrm r})^{\gamma}
\odot \referenceBracket, 1)
-\oneBracket
.
\]
\change{Here, we observe that the gamma function primarily scales the exposure ratio, leading to linearly scaled HDR values in the final output of our method. Since HDR reconstruction inherently suffers from a global scale ambiguity, this scaling does not pose a limitation. 
To further evaluate the impact of the CRF, we test a more complex model introduced by Eilertsen \etal \cite{eilertsen2017hdr}, defined as:}
\begin{equation}
\label{eq
}
CRF_{\beta,\gamma}(x) = \frac{(1+\beta) x^{\gamma}}{\beta + x^{\gamma}},
\end{equation} 
where $\beta \sim \mathcal{N}(0.6, 0.1)$ and $\gamma \sim \mathcal{N}(0.9, 0.1)$ represent the distributions of the CRF parameters derived from the analysis of a large dataset of real-world images \cite{wang2023glowgan}. We use the mean values of these parameters and re-run our method with this CRF model. The corresponding results, labeled as $\method{Ours} ^{\dag}$ in \refTab{Generation}, show no significant performance gains, suggesting that the simpler gamma model remains effective for our application.

\change{Based on these findings, we argue that the choice of CRF does not significantly affect the performance of our method}.

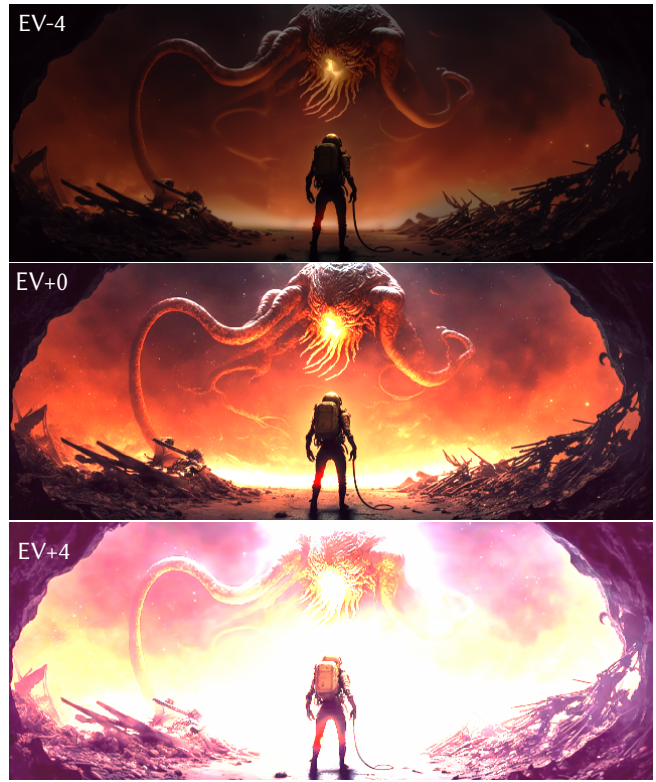
\myfigure{high_res}{Panoramic HDR generation at a $256 \times 640$ resolution given an AI-generated LDR image (middle row): To generate a panoramic image, we follow the diffusion composition technique from \cite{jimenez2023mixture} and simultaneously denoise three tiles of $256 \times 256$ resolution, each with a 64-pixel overlap, to ensure smooth transitions between them. The image-domain unconditional diffusion model \cite{dhariwal2021diffusion} serves as our base model for this process.}

\paragraph*{Extension to latent diffusion models}
The results presented so far are generated using the best-performing image-domain diffusion models. Although image-domain models have limited resolution, in \refFig{high_res}, we demonstrate that producing high resolutions with these models is still possible given enough computing time. However, to further enhance both the quality and resolution of image generation, we employ our DPS approach directly on latent diffusion models (LDMs) \cite{rombach2022high-resolution}, following the methodology outlined by Rout \etal \cite{rout2024solving}. In this context, we perform posterior sampling in the latent space, and accordingly, our prior and posterior scores in {\refEq{posterior_estimate}} are modified to:
\begin{equation}
    \label{eq:latentposterior_estimate}
    \nabla_{\bracketLatent_\iteration} \log p_\iteration(\bracketLatent_\iteration | \conditionsignal, \guidingsignal)
    \approx
    \scoreLatent(\bracketLatent_\iteration, \conditionsignal, \iteration)
    -
    \lambda 
    \nabla_{\bracketLatent_\iteration} \cost(\decoder(\posteriorBracketLatent_\iteration), \guidingsignal).
\end{equation} 
The rest of the equations, \refEq{costDown} and \refEq{costUp}, remain unchanged. Here, $\bracketLatent$ represents the latent code, $\scoreLatent$ is the score function of a pre-trained LDM, and $\decoder$ is the latent decoder that translates the latent code $\bracketLatent$ back into pixel space as $\bracket = \decoder(\bracketLatent)$. Note Rout \etal \cite{rout2024solving} also introduces a "gluing term" to penalize inconsistencies at mask boundaries; however, we did not find it necessary for our purposes.
In this experiment, we again apply the time-dependent $\lambda$ with $\lambda_0 = 2$ and perform T = 500 iterations to generate results. 
\refFig{text2HDRLatent} illustrates some examples for text-based generation at a resolution of 512$\times$512 using the pre-trained Stable Diffusion v-1.5 \cite{rombach2022high-resolution}.

\myfigure{text2HDRLatent}{Text-based HDR generation using the recent latent diffusion model \cite{rombach2022high-resolution} as the backbone. More examples are provided in our supplementary material.}

\paragraph*{Alternative solution to DPS}
We further investigate the alternative choice of score distillation sampling (SDS) \cite{poole2022dreamfusion} for HDR generation. The SDS method naturally allows for direct reconstruction of an HDR signal. In this approach, the optimized image can be represented by either a 2D-pixel grid or a neural network (NN); however, we found the NN provides better results than a simple pixel grid. During each optimization step, the HDR image is randomly exposed with EV+$x$, where $x$  is drawn from a normal distribution with a mean of zero and a standard deviation of four. We compute the SDS loss on the exposed images and update the parameters of the HDR image accordingly. The SDS loss guides the current estimate of the exposed images towards the manifold of natural images learned by the pre-trained diffusion model \cite{rombach2022high-resolution}. In \refFig{sds}, we present our best-effort results. While this simpler approach can generate HDR content, achieving natural results remains challenging.

\myfigure{sds}{HDR generation using SDS-based optimization \cite{poole2022dreamfusion}: the resulting images are HDR, but unfortunately not natural.}

\mysection{Limitations}{Limitation}
\change{Inheriting the properties of diffusion models, our proposed approach is inherently slow, especially compared to feed-forward methods like HDRCNN and MaskHDR ({\refTab{timing}}). This limitation is further exacerbated in our framework, as we simultaneously denoise multiple brackets, making it slower than the original DPS. The DPS framework typically requires a large number of diffusion steps to converge, significantly contributing to the slower sampling speed. Incorporating advanced sampling strategies, such as those proposed by Song \etal \cite{song2023pseudoinverse} and Zhu \etal \cite{zhu2023denoising}, can help address this bottleneck. Another constraint is the GPU memory requirement, which limits the number of exposure brackets that can be processed.}

\mysection{Conclusion}{Conclusion}
We have suggested a novel method for generating HDR images using a black-box diffusion-based image generation model without the need for expensive retraining or fine-tuning. The key idea is to generate multiple LDR brackets in a synchronized and consistent manner. Our approach is simple to implement, intuitive, and capable of producing results with unprecedented quality in the highlight regions while effectively reducing noise in shadows.
These capabilities have been validated through diverse applications of our method and comparisons with baseline techniques, demonstrating its effectiveness and versatility.
\change{Extending our approach to HDR video can be an interesting direction for future work, particularly in scenarios where EV+0 exposure varies across frames due to auto-exposure adjustments. This introduces challenges such as ensuring temporal consistency across frames. Additionally, other frame-specific factors, including motion blur, defocus blur, depth-of-field blur, and varying noise characteristics, will likely necessitate modifications to the proposed consistency terms. A particularly challenging task would be reconstructing an all-in-focus HDR frame from an input LDR image impacted by these distortions. Building on the consistency terms proposed in this work, similar strategies could also be employed to generate focal or depth-of-field stacks.}

\section*{Acknowledgment}
Open Access funding enabled and organized by Projekt DEAL.


\bibliographystyle{eg-alpha} 
\bibliography{paper}     

\end{document}


\teaser{
    \includegraphics[width=\textwidth]{figures/LDR2HDR_suppl}
	\caption{Additional LDR2HDR reconstruction results for LDR images generated by Stable-Diffusion. In this case, EV+0 is the input to our method.
    }
	\label{fig:Teaser}
}

\maketitle

\mycfigure{apline}{Text-based HDR generation: multiple generation results given the same input text prompt.}

\mycfigure{text_histogram_suppl}{Examples of text- and histogram-based HDR generation, where generated exposure brackets follow a text query while also matching the color histogram of a reference LDR image. On the left, reference LDR images are shown alongside their corresponding color histograms. In each row, we present results for different text prompts given the same input histogram. Alternatively, different rows showcase the results of different histograms with potentially similar prompts.  }

\mycfigure{text2HDR_suppl}{More examples of text-based HDR generation.}

\mycfigure{text2HDR_latent_suppl}{Additional examples on text-based HDR generation using latent diffusion model \cite{rombach2022high-resolution} as the backbone.}

\clearpage

\bibliographystyle{eg-alpha} 
\bibliography{paper}   